# Reviewers' ratings and bibliometric indicators: hand in hand when assessing over research proposals?

Álvaro Cabezas-Clavijo[1], Nicolás Robinson-García[1]*, Manuel Escabias[2], Evaristo Jiménez-Contreras[1]

[1] EC3: Evaluación de la Ciencia y de la Comunicación Científica, Facultad de Comunicación y Documentación, Universidad de Granada, Spain.
Email: acabezasclavijo@gmail.com; {elrobin, evaristo}@ugr.es
[2] Departamento de Estadística e Investigación Operativa, Facultad de Comunicación y Documentación, Universidad de Granada, Spain.
Email: escabias@ugr.es
* To whom all correspondence should be addressed

**Abstract**

*Background:* The peer review system has been traditionally challenged due to its many limitations especially for allocating funding. Bibliometric indicators may well present themselves as a complement.

*Objective:* We analyze the relationship between peers' ratings and bibliometric indicators for Spanish researchers in the 2007 National R&D Plan for 23 research fields.

*Methods and materials:* We analyze peers' ratings for 2333 applications. We also gathered principal investigators' research output and impact and studied the differences between accepted and rejected applications. We used the Web of Science database and focused on the 2002-2006 period. First, we analyzed the distribution of granted and rejected proposals considering a given set of bibliometric indicators to test if there are significant differences. Then, we applied a multiple logistic regression analysis to determine if bibliometric indicators can explain by themselves the concession of grant proposals.

*Results:* 63.4% of the applications were funded. Bibliometric indicators for accepted proposals showed a better previous performance than for those rejected; however the correlation between peer review and bibliometric indicators is very heterogeneous among most areas. The logistic regression analysis showed that the main bibliometric indicators that explain the granting of research proposals in most cases are the output (number of published articles) and the number of papers published in journals that belong to the first quartile ranking of the Journal Citations Report.

*Discussion:* Bibliometric indicators predict the concession of grant proposals at least as well as peer ratings. Social Sciences and Education are the only areas where no relation was found, although this may be due to the limitations of the Web of Science's coverage. These findings encourage the use of bibliometric indicators as a complement to peer review in most of the analyzed areas.

**Keywords**
Bibliometric indicators, peer review, Spain, grant proposals, research funding, research policy, evaluation agencies

# Introduction

A key issue regarding research policy has to do with the allocation of funds. The most extended system for doing so is peer review. However, one of the traditional debates in research evaluation has to do with its reliability. Although it is considered the most effective system, peer review has been long criticized by the community, stating that it propitiates endogamy and a closed-minded growth of science [1-2]. It is perceived as a kind of black box in which it is not really clear what peers conceive as quality and which aspects are considered as key factors for success. Many studies have been made devoted to the analysis and validation of peer review [1, 3-6], but none has been able to establish sound conclusions on this regard. Their main limitations are the lack of large data sets and no consensus whereas to the interpretation of results [7].

These concerns along with others such as the inconsistency, slowness, potential biases and high costs of peer review [3], or the subjectivity and heterogeneity of reviewers [8] have led funding agencies and researchers to focus on bibliometric indicators as they can offer quantitative measures that appear much more reliable and easier to use when quantifying the results of the investment made in science [9]. This line of thought follows a generalized and reasonable perception considering that bibliometric indicators should go in accordance with peers' judgment to some extent, as they are supposed to measure similar attributes. As a consequence, research policy-makers' interest on transforming national research systems into competitive entities has led to the inclusion of bibliometric indicators in their assessment systems, in some cases along with peer review [10] or just exclusively [4, 11]; enabling mechanisms that can lead them to monitor and distribute research funding at an institutional level.

Although bibliometric indicators seem to work reasonably well at national and institutional level [4], concerns arise when applied at an individual level. According to Allen and colleagues [12], there is correlation between expert opinion and performance, as measured by bibliometric indicators, but a sole reliance on bibliometrics may omit papers containing important results which would be considered by expert review. Notwithstanding this limitation, bibliometric indicators are frequently used by decision-makers and science policy managers who are urged to support their decisions with proof [13]. To this end, many indicators have arisen in order to synthesize both the qualitative and quantitative dimensions of research, being the h-index and its many variants the most popular bibliometric indicators aimed at evaluating individuals [14].

However, no matter the validity of such indicators, many countries still rely heavily on journal rankings [11] as a proxy for research quality. In this sense, it is also usual to assign impact factors of journals to individual papers as proxy of their impact, even if it is proved to be an erroneous practice, given the skewness of the citation distribution of publications [15]. Consequently, most studies conclude that citation analysis and bibliometric indicators could be used when taking into account decisions regarding research funding, especially for the hard sciences [16]; but never as a substitute for the peer review system and simply as a complementary tool. This approach is known as "informed peer review" [17]. The idea is to create useful products that are based on bibliometric methods, easy to understand that can be used by reviewers to orient their assessment, or by funding agencies in order to monitor and control researchers' strengths and weaknesses.

Following this line of thought, one may consider bibliometric indicators as a possible solution to minimize the shortcomings of peer review. Many studies can be found in the literature analyzing the success in different countries which include bibliometric indicators within their national research systems for allocating funds [10, 18-22]. This study presents further evidence on the relation bibliometric indicators and peer review and their level of coincidence when predicting research funding decisions. However, most of these studies normally focus on few research areas; in this case we present evidence for 23 different fields which cover all of the research areas except for those from the Arts & Humanities area. We focus on the Spanish case which follows a similar funding system to that of many other countries; allocating funds for grant applications according to the contents of the research project and to the recent past performance of the Principal Investigator (hereafter PI) and their research team. In summary, Spanish research funds are distributed through four main channels [23]: (1) a human resources selection system based on position status associated with salary; (2) a competitive project-funding system divided into different programs; (3) a reward system based on credit and reputation; and (4) other channels based on contractual agreements or private funding.

This paper is focused on the second channel, that is, the main system for research funding. In this sense, our main goal is to measure the relation between ratings assigned by reviewers when assessing grant proposals and bibliometric indicators derived from PIs' previous research performance. The study will be mainly focused on the PIs' curricula, assuming that the approval of funding applications relies heavily on their CV and that researchers with high ratings will also perform well when applying bibliometric indicators. This is the first study of such characteristics analyzing the Spanish research funding system. Parting from these main objectives, we try to determine the bibliometric factors that influence the final decision for funding a research project. For this, we pose the following research questions (RQ).

RQ1. To what extent do peer review ratings of grant proposals predict the funding decisions, in total, and differently across scientific areas? Are PIs' curricula determinants on the concession of a research grant?

RQ2. Are bibliometric indicators influential? Which (if any) increase the chances of being funded?

## Material and methods

Our main goal is to study the relationship between ratings assigned by peer review to grant applications and bibliometric indicators of past research performance for their PIs, as well as the predictability of these indicators for granting research projects. In this section we present an overview on the peer review process and the data processing and calculation of the bibliometric indicators. For this, we will first describe the population of researchers analyzed, the indications reviewers follow, the process for evaluating grant applications and how is the final decision taken (concession or rejection of the research proposal). Then, we define the bibliometric indicators used, data collection and processing, and the statistical analyses undertaken.

### The peer review process: Research evaluation in Spain

The grant proposals system in Spain is monitored mainly but not exclusively, by the National Agency for Evaluation and Foresight (hereafter ANEP, Spanish acronym) through the National R&D Plans. It should be noted that criteria used by this agency has been much influenced by the patterns followed in the Basic Sciences, as researchers from these fields greatly supported the creation of the first evaluation agencies during the 1980s [24]. Hence, Thomson Reuters Web of Science and its derived products, especially the Journal Citation Reports (hereafter JCR), are considered a keystone of research funding and rewarding in most research fields playing an overriding role for the internationalization of Spanish research and the adoption of international standards [25]. Despite criticisms to the JCR impact factors [26, 27], this indicator has been used greatly in Spain. The National R&D Plans are the most important research grant system for funding research projects in this country. These projects last 3 years and are led by a researcher who is considered fully responsible for the execution of the project. They provide the Spanish research system with its main channel of funding, enabling it to develop research policies, transparency in the distribution of funding and the inclusion of a set of international standards and good practices among researchers.

The Plans are assessed by the ANEP, which is in charge of the *ex ante* assessment of applications and their applicants by means of peer review. After that, grant proposals scores are sent out to the Minister responsible for research policy, which has the final decision over the fate of the applications.

In the present study we focus on the 2007 call. In Figure 1 we show the process followed for the evaluation of grant applications. We analyzed the total population of applications of individual projects sent to type B, that is, a total of 2333 applicants, which represent 82.03% of the whole share of applications to the R&D National Plan. It is important to note that the candidates were not allowed to lead more than one project at the same time within the R&D Plan framework; therefore there is only one application per candidate. Data of the PI (name and affiliation) and research area were provided by ANEP. After the evaluation process ended, this agency supplied a second list with the scores assigned by the reviewers for each section. Each project proposal is assessed by two reviewers chosen by the coordinator of the specific research area, giving a score to each of the assessed criteria [28], all of which are highly subjective as no clear definitions are provided. These criteria are based on five sections where the highest score means excellent: principal investigator's curriculum (16-point rating scale), research team's curricula (10-point rating scale), goals (8-point rating scale), relevance (8-point rating scale) and viability of the proposed research project (8-point rating scale). Although two referees evaluate each proposal, the agency provides one final rating for each proposal which is assigned by the coordinator according to the referees' reports. In this sense, ANEP states that there are high levels of agreement between referees' ratings. Finally, data with all the accepted proposals was downloaded from the Ministry of Science website.

**[Figure 1.]**

A total of 2333 type B grant applications for individual projects were received for the 2007 National R&D Plan. From these, 1479 (63.4%) were finally accepted and funded (Table 1). The areas with a highest number of proposals accepted were Fundamental & System Biology, with 232, Chemistry with 132 and Physics with 103, on the other hand, Clinical Medicine (7 proposals accepted), Civil Engineering & Architecture (18) and Education (38) were the areas with the lowest number of proposals accepted. In relative terms, differences are also important. The area with the highest success rate was Physics with 83.1% of its applications accepted, followed by Mathematics (79%) and Chemical Technology (77.3%). Applications from Biomedicine, Social Sciences, Economy, Education, Civil Engineering & Architecture, Clinical Medicine and Psychology had more than half of their proposals rejected, with Clinical Medicine (21.9%), Education (40.9%) and Biomedicine (41.9%) being the three areas with the lowest success rates.

**[Table 1]**

## Data processing, bibliometric indicators and statistical analyses

In order to test the relation between bibliometric indicators and peer review we selected a five-year period prior to the research funding call (2002-2006) which is the period reviewers must evaluate according to the funding call when assessing on the candidates' research performance. Then, we downloaded applicants' output from the Thomson Reuters Web of Science database between February 2009 and May 2010. Citations for every paper were also retrieved, restricting the citation window from 2002 to 2008. This citation window was selected in order to allow the most recent publications to be cited. The search was conducted manually, one-by-one, taking into account possible name variations and affiliation changes during the study period. The following document types were analyzed: articles, reviews, letters, editorial material and proceedings papers. This data was introduced in a relational database along with information provided by the ANEP (names, project code, type of project, affiliation, score ratings, papers published by PIs during the study period, concession of the project and funding received). Also, journals' impact factors were downloaded from the JCR. This way we can relate journals in which PIs published with their Impact Factor in the same publication year and hence, identify first quartile papers (for a detailed explanation of the considered variables see Table 2).

**[Table 2]**

In this context, output should be interpreted as a quantitative measure for international outcome of the PI, while Q1 and %Q1 must be considered not only as visibility indicators, but as proxies to measure the prestige of journals and hence the authors' competitiveness. By the same token, citations are understood to be a valid measure of the impact of PI's research. Although the latter dimensions of research are related (visibility and impact) as publications in high impact journals tend to gather more citations than papers in low impact journals, both could influence separately or jointly on reviewers' judgment. However both have been considered in the discussion as qualitative measures. The conclusions derived from this study are supported by various statistical methods and analyses. Although the main results are present in this paper, we have also included Complementary Material (available at http://hdl.handle.net/10481/23451) in order to enrich the analysis and provide the reader with further information.

Although it is obvious that the final decision on the granting of research proposals depends on the ratings assigned to the five sections analyzed by reviewers (PIs CV, Research Team' curricula, objectives, relevance and viability), the importance given by reviewers to each section may vary among areas. For this reason we decided to fit a logistic regression model to analyze if the concession of grant proposals can be determined from the ratings of each section and for each area. The selection of the most important sections and the order by which they are considered in the model were undertaken by means of a stepwise regression. These results are shown in table 1 (Complementary material). From such fit we derive that the model can predict correctly around 90% of the cases based on the area under the ROC curve. In this study we consider that the concession of grant proposals is determined by the past research performance of the PI. In order to prove if this premise is correct we compared the results of each fit for the various logistic regressions with those obtained if the only covariable was PI's ratings. In Table 3 we show the area under the ROC curve, the Correct Classification Rate and the $R^2$ coefficients.

In order to compare the distributions between granted proposals and rejected proposals for each of the considered bibliometric indicators, we obtained box plot diagrams (see Complementary Material, Figures 1-11). Such diagrams clearly show the differences between the distributions. However, we tested the statistical significance of such differences by means of a Wilcoxon signed-rank test (Table 5). We chose the Wilcoxon signed rank test [29-31] due to the skewness of the distribution of most variables [15]. It was performed one-sided as in most areas the median values of the bibliometric indicators are lower for rejected proposals than for accepted proposals (see Table 4).

Then, as referees' ratings are not strictly a continuous variable, we used the Spearman and Pearson coefficients in order to see if there is any association between each of the aspects assessed by referees in all areas (Table 6). Next, we performed a stepwise linear regression analysis [32] in order to select the bibliometric variables that can better explain the ratings assigned to the PI of each project for each of area (Complementary Material, Table 2). Finally, as the results were not satisfactory, we performed a multiple logistic regression analysis [33-35] in order to explain the granting of research proposals (probability of acceptance) by using bibliometric variables to each of the areas analyzed. We used the stepwise analysis to determine which variables and in which order they better explain the granting or rejection of research proposals (Table 7). The results of such analysis would allow us to see if the use of bibliometric indicators would be enough to predict the concession of research proposals and therefore, substitute the peer review process. Also, this model identifies for each area which variable has more importance on the prediction of the acceptance of projects and how it influences it. The software programs used for such analyses were XLStat 2009 3.02 and R 2.14.1.

## Results

### Description of referees' ratings, bibliometric indicators and granted vs. rejected distribution of grant proposals

In table 3 we show the area under the ROC curve (hereafter AUC), $R^2$ and the Correct Classification Rate (hereafter CCR) for two possible scenarios on the variables which better explain the concession of grants according to the reviewers' ratings. As observed, when introducing only the ratings for each section assessed, AUC and $R^2$ as well as CCR are very similar to what happens when we only introduce PI's ratings as an explanatory variable. These results allow us to assume that when PI's are favorably rated they have more probabilities of having their grant applications approved.

**[Table 3]**

Table 4 shows the median values for OUTPUT, AV CITATIONS and %Q1 of granted vs. rejected grant proposals. Also, it shows the median values for the referees' ratings of the grant applications. This way the reader can observe differences between the bibliometric performance of applicants and the final score their applications received. When only considering researchers of proposals accepted, Chemistry (21.5) was the area with the highest median scientific output, along with Biomedicine (19.5). Among the proposals rejected, Clinical Medicine had the highest median output with 9 papers per researcher. Education was the only field that did not follow this pattern. The median value of citations per paper was 6. This indicator doubles for proposals accepted (7.8) when compared with proposals rejected (3). In only one area the median value was the same for accepted and rejected proposals (Education). Scientific output published in Q1 journals was 37.5%, with significant differences between proposals accepted (50.0%) and proposals rejected (16.7%).

If we consider the PIs' curricula, it is striking that proposals rejected from areas such as Vegetal & Animal Biology / Ecology or Social Sciences reach maximum ratings that equal proposals accepted (15 for the former, 16 for the latter). This behavior is also found in other areas, for instance, proposals in Mathematics and Physics where PIs' CV had low ratings (5 out of 16) were finally funded.

**[Table 4]**

In order to test if the differences between medians of the bibliometric indicators of PIs' CVs for granted and rejected proposals were significant; in Table 5 we show the results after applying a Wilcoxon signed-rank test. We show the Wilcoxon-test value (z) and ρ-values for each indicator. In bold, we highlight the ρ-values of bibliometric indicators and areas in which significant differences were found. In 14 of the 23 areas under study, there were statistically significant differences between the values of all bibliometric indicators for granted and rejected proposals. As observed, Education was the only field for which none were found for any of the bibliometric indicators. Computer Science & Information Technology and Social Sciences showed differences for only two of the five indicators analyzed (AV CITATIONS, &Q1 and Q1 for the former and OUTPUT, %Q1 and Q1 for the latter). The two indicators which showed less differences were AV CITATIONS and %Q1 (in both cases differences were not significant for five areas).

**[Table 5]**

## Influence of bibliometric indicators on peers' ratings

At this stage, it is interesting to study if bibliometric indicators could be used as predictors of the referees' ratings and if they go hand in hand with their judgments. For this, as a prior step we analyze the correlation between the PIs' CV scores each application received and the bibliometric indicators selected (Table 6). Due to the differences of the nature of peers' ratings and bibliometric indicators, we used both, Spearman and Pearson coefficients. In general terms, the correlation is very heterogeneous with very low or zero correlations on the one hand, and from moderate to high correlations (0.50-0.75) on the other. When using the Pearson coefficient, no area or indicator seems to correlate significantly with the ratings assigned by the referees. However, when using the Spearman coefficient, correlations are slightly higher. In fact, there seems to be some correlation (Spearman $\geq 0.70$) in two areas; Electrical, Electronic & Control Engineering and Mechanical, Naval & Aeronautic Engineering. Although in each case, the indicators are different. While the first shows correlation between OUTPUT and referees' scores (0.73), the latter shows correlation for CITATIONS and Q1 (0.75 and 0.72). On the other end we find Education, in which not only ratings and bibliometric indicators are independent, but even in some cases correlations are negative. The other area with values near to zero is Social Sciences.

**[Table 6]**

Despite the scarce correlation between each bibliometric indicator, we could still assume that jointly, these indicators influence or at least explain reviewers' ratings when evaluating the PI's CV. In order to test such hypothesis, in Table 2, Complementary Material, we performed a linear regression analysis, selecting the variables that best explain the model through a stepwise method. However, results were not satisfactory and ruled out this possibility as concluded from the values of the coefficient of determination. Nevertheless, we considered that these results did not rule out our hypothesis and used a different approach.

In Table 7 we apply a logistic regression analysis stepwise by area, in order to see if the bibliometric indicators could explain the final decision taken for granting or rejecting research proposals. For each area, we show the variables selected by the stepwise method, z and ρ-values of the goodness of fit to the logistic model, that is, the test which indicates if the logistic model is adequate or not for modeling the concession or rejection of grants. Next, we show some precision measures on the predictions made, such as AUC and the CCR. Finally, the odds ratio of each explanatory variable is included, in order to explain the relation between the indicator and the final concession or rejection of the grant application. The odds ratio is a value that multiplies the advantage of obtaining a research grant in opposition to having the applications rejected for each unit of a given indicator. Therefore we observe that the AUC ranges from 0.73 to 0.89 and only in one case (Education) it shows lower values. Also CCRs are very high and only in Education it shows lower figures than 60%, reaching the highest values in the areas of Biomedicine and Clinical Medicine & Epidemiology (82.56% and 81.25% correspondingly). When observing the variables that better explain the granting of research proposals, OUTPUT seemed to be the variable which affected the most areas in first place (10), followed by Q1 (9). CITATIONS and AV CITATIONS only positioned themselves as explanatory in first place in two areas; Economy and Livestock Farming & Fishery. On the rest of these areas these variables are present but always on second or third place.

[Table 7]

Finally, we include the Intercept value which indicates the odds of receiving a research grant versus having a rejected application. The number in brackets shows the odds of receiving the research grant versus having it rejected. For example, in the case of Agriculture, the Intercept value is 0.36, which means that a PI with Q1 publications has a probability 2.74 times higher of receiving a research grant than the one who has no Q1 publications.

## Discussion and concluding remarks

Before discussing the results of this study, it is necessary to acknowledge several shortcomings that affect the work. Firstly, the population of researchers in some areas is not enough to generalize these results. Particularly, results on areas such Clinical Medicine & Epidemiology and Civil Engineering & Architecture are based on less than 50 individuals. This calls for caution when interpreting the results obtained. Another limitation has to do with the methodology employed as the database selected is considered to have a limited coverage for Social Sciences and Engineering [36]. This limitation mainly affects three of the areas assessed (Civil Engineering & Architecture, Social Sciences, Education Science), in which more than a third of the population does not have papers indexed in this database. The other two areas within Social Sciences (Psychology and Economy) range from 13% to 17% of the individuals with no production in this database, while in all the other areas this percentage drops below 10%. The reason for using this database and not considering other sources has to do with its high reputation among funding agencies as a reflection of international contributions. Spanish scientific policy has been directed towards the internationalization of researcher's output; meaning publishing in JCR journals including those areas which are considered to not be well covered by this database such as Engineering and Social Sciences. Finally, another shortcoming that mainly affects areas from the Social Sciences is the type of document considered. Books and book chapters, which play an important role in these areas have not been considered in this study, despite the fact that these publications are also evaluated by reviewers, along with other aspects of researchers' curricula which are also considered to be part of their research activity such as, leadership in other research projects, number of dissertations supervised, or when referring to the Applied Sciences, the number of contracts signed with firms or of patents registered.

The present study analyzes the relation between peer judgment and bibliometric indicators, and how these indicators affect the applicants' chances for being funded. For this, we studied the population of researchers (n=2333) who applied for a grant proposal in the main call for funding within the 2007 Spanish R&D Plan. We analyzed the relationship between reviewers' ratings and bibliometric indicators for the 2002-2006 time period. The suggested hypothesis was that peer judgment would correlate highly with bibliometric indicators. For this, two research questions were posed.

RQ1 To what do peer review ratings of grant proposals predict the funding decisions, in total, and differently across scientific areas? Are PIs' curricula determinants on the concession of a research grant?

Concerning this question, the significant differences found in most of the areas suggest that grant proposals are usually conceded as a function of the PI's research performance (Table 3), which is a key factor in the final decision. This is understandable as these funding programs tend to assume that researchers with a solid background may ensure the future success of funded research. Such premise is based on the lack of ex-post evaluation on the fate of the funded proposals. As pointed out by Sanz-Menéndez [37], the peer review process based on past performance implicitly assesses on the future performance of the proposal. It also indicates that peers are predisposed to rate positively researchers with a well-established background regardless the contents of their project. There is an heterogeneous correlation between reviewer ratings and bibliometric indicators, although results suggest the latter influence reviewers' behavior when assessing grant proposals. This perceived influence is noted in Table 4 and Figures 1-5 (Complementary Material) where performance is significantly lower for the curricula of applicants' with proposals rejected. Mechanical, Naval & Aeronautic Engineering and Electrical, Electronic & Control Engineering showed a more consistent correlation between bibliometric indicators and curricula ratings when using the Spearman coefficient. However, we cannot state that reviewers in these areas take into greater consideration bibliometric criteria than in others. These differences in the correlation between curricula ratings and bibliometric indicators may be due to the shift from a qualitative scale (reviewer opinion) to a quantitative scale (reviewer rating), that may blur this relation.

Another aspect that may affect this lack of correlation may be the amplitude of the rating scale (from 1 to 16 for curricula) which does not go in accord with bibliometric indicators that can potentially range from zero to the infinity. This reduces inevitably the ratings to a much limited scale, minimizing differences among applicants. Therefore, the difference in the average number of publications for researchers whose projects were accepted is of 110% comparing with rejected proposals. Regarding the average number of citations it is of 93%. When focusing on ratings, the differences are just of 42%. Also, different biases, for instance the reviewers' predisposition to evaluating positively (Table 4) or those described by Wessely [6] may affect this final score. In Spain, the fact that reviewers are highly experienced researchers may favor the agreeableness of the evaluations due to the small size of the national research system and the invisible colleges that surround it.

We can deduce from these results that the two Social Science areas (Education and Social Sciences and Economy) have low correlations between bibliometric indicators and curricula ratings (Table 5). The fact that these areas were not well represented in the Web of Science database for the publication period assessed (up to 2006), might condition the importance reviewers assign to it. The lack of predictability between bibliometric indicators for proposals accepted and proposals rejected in certain areas such as Education or Social Science (Table 6) lead us to believe that the criteria used by reviewers are not homogenous. The main reason for this may be the importance of national publications and other types of publications. This is supported by the fact that these areas show (with Civil Engineering & Architecture) the highest percentage of proposals accepted for which the PI has no WoS publications during the study period (47.4% in Education; 31.4% in Social Sciences). In the case of Education, it is even more remarkable, as the percentage of proposals by researchers with no publications in the WoS database and funded is even higher than the rate of non-productive researchers found in the sample studied.

Even so, this is a peculiar fact, as in the last decade, Spanish research policy has been directed towards favoring international publications, changing Spanish researchers' habits and causing a migration from national journals to international ones (meaning international those journals indexed by Web of Science) [38]. Evaluators may also be considering other types of documents not reflected in our study such as national journals, books or book chapters. The high percentage of non-productive researchers in Education and Social Sciences suggests the need for further research using additional information sources such as the recently launched Book Citation Index [39], and national or regional databases. In fact, many of these alternative databases are already used in some research assessment exercises at a micro-level.

## RQ2 Are bibliometric indicators influential? Which (if any) increase the chances of being funded?

The indicators that most influence research granting among the studied variables are OUTPUT and Q1 publications. Differences are found within fields. Those belonging to Engineering & Technology are the ones in which bibliometric indicators seemed to better explain the final granting decision (Table 6). Also, we found that, despite the shortcomings above discussed regarding the areas of Education and Social Sciences, research impact (considered as Q1 publications and number of citations) work as influential indicators in the chances of being funded for the other two areas of the Social Sciences; Economy and Psychology. These two fields have shifted towards an internationalized research context and therefore, the Web of Science seems to be a good bibliometric resource for analyzing the Spanish research activity in these fields.

Generally speaking, reviewers value better the quantity of research output (considered as such publications indexed in WoS) than its quality (considering as such papers published in Q1 journals) in technology and engineering areas, as well as in some basic areas like Mathematics or Physics. Impact and visibility appears to be more important than the size of the PI's recent output in biological and biomedical fields as well as for Agriculture and Livestock Farming and Fishery. At this point it is important to emphasize that ANEP does not decide whether a proposal must be accepted or rejected, but assess only on the proposals and, afterwards, an experts panel selected by the Ministry of Science takes the ultimate decision according to the reviewers' reports and other political criteria. Amongst them there is for instance, a priority over strategic research fronts or gender or geographical criteria. These factors have not been studied in the present paper, however, they have a marginal effect on the final decision as observed in Table 3 where CCR for total ratings show figures above 0.80 for all areas except three of them and always above 0.70. However, findings in this study suggest that the bibliometric indicators applied to the PI's publications in WoS influence to a great extent in most of the studied areas (except Education and Social Sciences) the fate of a proposal, emphasizing its success on explaining the concession for research funding in Basic and Health Sciences and to a lesser extent in other areas closer to the Social and Applied Sciences (Psychology, Food Science & Technology, Computer Science & Information Technology).

The results show low correlation between bibliometric indicators and reviewers ratings (Table 5). However, we must take into account that other factors different than those reflected in this study may also influence on the final rating of the PIs' curricula, such as their leadership in research projects, number of supervised theses, or as in the case of social sciences, publication of monographs or book chapters. However, bibliometric indicators explain reasonably well the final decision on granting research proposals (Table 7) and thus, we suggest they could be used as a complement to the peer review process when assessing researchers' curricula, as long as the criteria used fits to each area. Indeed, it seems that peer review and bibliometric indicators are not fully independent and that reviewers use bibliometric raw data when assessing researchers' curricula. If so, one could consider that such evaluation could be complemented with bibliometric indicators. For instance, with the construction of reference thresholds that can help experts when comparing applicants' previous performance with the general performance of researchers in the same area of expertise, as has happened in Spain [40]. Evidences from Italy, a country with a very similar research system, suggest that, at least for the Sciences area, the peer review system does not pay off when assessing researcher's output as results don't differ substantially from those obtained by bibliometric means [41]. From the findings of this study, we also suggest the encouragement of indicators that emphasize the quality of research output (publications in Q1 journals, the h-index or the average of citations per paper) rather than quantity, as researchers tend to match assessment criteria [10, 25]. This way, peers judgment would only be used to assess the content of scientific proposals.

Evaluation processes are complex and arouse controversy, as happens with the British Research Excellence Framework in which, after several studies and surveys, the number of citations will only be used when assessing as a bibliometric tool to complement expert judgment in a limited number of areas. However, in the Spanish case, where bibliometric assessment has become usual, we believe that the establishment of a system similar to that developed in the UK would not raise the same reactions. Since the 1980s, the Spanish research system has experienced a great increase on its institutional size and in its capacity to produce quality research, complying with international standards. In this sense, the evaluation processes undertaken by ANEP have fulfilled their mission reasonably well, contributing to the improvement of Spanish research. However, the current economic context dominated by cuts in R&D and the restructuring in universities aimed at increasing the quality of research and making a more efficient system, may end with the current R&D funding and assessment systems in Spain. In this context, research evaluation processes are more relevant than ever and must be conducted with the greatest precision and reliability, modifying and adapting them if necessary in order to improve the efficiency of the system.

This paper focuses on the relation of bibliometric indicators and peer review and the level of concordance between each other. This is a topic of great importance to managers and research policy makers as bibliometric indicators are more economically viable and seem to be more objective than peer review judgment. From our findings we conclude that there isn't seem to be a direct relation between bibliometric indicators and experts' ratings, however they both lead to the similar results when deciding on the granting of research proposals.

### 7. Acknowledgements
The authors would like to thank Rodrigo Costas and Antonio Callaba de Roa for their helpful comments in previous version of this paper as well as the two anonymous reviewers for the constructive comments. We would also like to thank Bryan J. Robinson for revising the text. Nicolas Robinson-García is currently supported with a FPU grant from the Spanish government, Ministerio de Economía y Competitividad.

**Figure legends**

**Figure 1.** Flowchart of the evaluation process of grant applications for the 2007 Spanish R&D Plan.

**Types of applications:** Type A is devoted for young researchers; Type B is intended for all researchers; Type C is devoted to research projects which need extraordinary sums of funding. **Types of projects:** Individual projects are led by a PI; coordinated projects imply several research groups with a coordinator and 2 or more PIs who apply separately in different applications.

# Tables

**Table 1.** Areas, total applications and applications granted per area

| ACRONYM | AREA | APPLICATIONS | GRANTED | % GRANTED |
|---|---|---|---|---|
| FSB | FUNDAMENTAL & SYSTEM BIOLOGY | 314 | 232 | 73.9 |
| CHE | CHEMISTRY | 187 | 132 | 70.6 |
| VAB | VEGETAL & ANIMAL BIOLOGY / ECOLOGY | 126 | 83 | 65.9 |
| PHY | PHYSICS & SPACE SCIENCES | 124 | 103 | 83.1 |
| PPH | PHYSIOLOGY & PHARMACOLOGY | 118 | 82 | 69.5 |
| ECO | ECONOMY | 117 | 57 | 48.7 |
| PSY | PSYCHOLOGY | 113 | 54 | 47.8 |
| SSC | SOCIAL SCIENCES | 108 | 51 | 47.2 |
| MST | MATERIALS SCIENCE & TECHNOLOGY | 107 | 77 | 72 |
| MTM | MATHEMATICS | 105 | 83 | 79 |
| ESC | EARTH SCIENCES | 97 | 67 | 69.1 |
| EDU | EDUCATION SCIENCE | 93 | 38 | 40.9 |
| FST | FOOD SCIENCE & TECHNOLOGY | 90 | 54 | 60 |
| AGR | AGRICULTURE | 86 | 47 | 54.7 |
| BMED | BIOMEDICINE | 86 | 36 | 41.9 |
| CSI | COMPUTER SCIENCE & INFORMATION TECHNOLOGY | 80 | 46 | 57.5 |
| CHT | CHEMICAL TECHNOLOGY | 75 | 58 | 77.3 |
| ECT | ELECTRONIC & COMMUNICATION TECHNOLOGY | 72 | 48 | 66.7 |
| LFF | LIVESTOCK FARMING & FISHERY | 59 | 35 | 59.3 |
| EEC | ELECTRICAL, ELECTRONIC & CONTROL ENGINEERING | 57 | 38 | 66.7 |
| MNA | MECHANICAL, NAVAL & AERONAUTIC ENGINEERING | 50 | 33 | 66 |
| CEA | CIVIL ENGINEERING & ARCHITECTURE | 37 | 18 | 48.6 |
| CLIM | CLINICAL MEDICINE & EPIDEMIOLOGY | 32 | 7 | 21.9 |
| **TOTAL** | | **2333** | **1479** | **63.4** |

**Table 2.** Description of the indicators used in this study.

| | Indicator | Definition | Acronym |
|---|---|---|---|
| **BIBLIOMETRIC INDICATORS** | Research output | Publications by PI and research field for the 2002-2006 time period | OUTPUT |
| | First quartile papers | Output in journals listed as first quartile (top 25%) in their JCR Subject Category when sorted by their Impact Factor by PI and research field for the 2002-2006 time period | Q1 |
| | Percentage of first quartile papers | Percentage of the output in journals from the 1st quartile of their JCR Subject Category by PI and research field for the 2002-2006 time period | %Q1 |
| | Citations received | Total of citations received by PI and research field for the 2002-2006 time period | CITATIONS |
| | Average of citations | Average of citations received by PI and publication and research field for the 2002-2006 time period | AV CITATIONS |
| **PEERS' CRITERIA** | PI's curriculum | Peers' judgment on the PI's research performance for the 2002-2006 time period | PI |
| | Research team' CV | Peers' judgment on the research team's research performance for the 2002-2006 time period | RESEARCH TEAM |
| | Goals of the research project* | | GOALS |
| | Relevance of the research project* | | RELEVANCE |
| | Viability of the research project* | | VIABILITY |

* These variables are not defined explicitly by the ANEP.

**Table 3.** Prediction ability measures of the logistic regression analysis to model the concession of research grants. First three columns considering as covariates the different sections' evaluated by reviewers and selected by the stepwise method. Last three columns only with PIs' ratings as covariate

| AREA | Ratings for each section | | | Ratings for Pis'CV | | |
|---|---|---|---|---|---|---|
| | AUC | $R^2$ | CCR | AUC | $R^2$ | CCR |
| AGR | 0.93 | 0.68 | 0.88 | 0.87 | 0.50 | 0.79 |
| BMED | 0.95 | 0.73 | 0.85 | 0.92 | 0.66 | 0.81 |
| CEA | 0.94 | 0.75 | 0.92 | 0.87 | 0.51 | 0.76 |
| CHE | 0.96 | 0.75 | 0.89 | 0.90 | 0.54 | 0.82 |
| CHT | 0.96 | 0.76 | 0.91 | 0.86 | 0.49 | 0.81 |
| CLIM | * | * | * | * | * | * |
| CSI | 0.95 | 0.72 | 0.88 | 0.86 | 0.50 | 0.76 |
| ECO | 0.98 | 0.86 | 0.93 | 0.95 | 0.76 | 0.87 |
| ECT | 0.87 | 0.55 | 0.79 | 0.86 | 0.48 | 0.76 |
| EDU | 0.91 | 0.61 | 0.81 | 0.84 | 0.45 | 0.75 |
| EEC | 0.98 | 0.84 | 0.93 | 0.88 | 0.54 | 0.79 |
| ESC | 0.89 | 0.56 | 0.80 | 0.82 | 0.39 | 0.76 |
| FSB | 0.95 | 0.71 | 0.88 | 0.91 | 0.60 | 0.82 |
| FST | 0.96 | 0.77 | 0.86 | 0.87 | 0.52 | 0.84 |
| LFF | 0.88 | 0.56 | 0.78 | 0.78 | 0.36 | 0.69 |
| MNA | 0.93 | 0.63 | 0.84 | 0.89 | 0.53 | 0.84 |
| MST | 0.93 | 0.65 | 0.86 | 0.84 | 0.41 | 0.75 |
| MTM | 0.96 | 0.73 | 0.90 | 0.91 | 0.57 | 0.81 |
| PHY | 0.87 | 0.41 | 0.83 | 0.83 | 0.32 | 0.83 |
| PPH | 0.93 | 0.61 | 0.84 | 0.90 | 0.56 | 0.83 |
| PSY | 0.95 | 0.73 | 0.89 | 0.89 | 0.56 | 0.81 |
| SSC | 0.91 | 0.60 | 0.83 | 0.82 | 0.39 | 0.75 |
| VAB | 0.94 | 0.72 | 0.90 | 0.86 | 0.50 | 0.80 |

* The logistic model does not apply to the data

**Table 4.** Median values for PIs' output, citations, Q1 publications and ratings indicators per area

|  | OUTPUT | | AV CITATIONS | | %Q1 | | PEERS' RATINGS | |
| --- | --- | --- | --- | --- | --- | --- | --- | --- |
| AREA | GRANTED | REJECTED | GRANTED | REJECTED | GRANTED | REJECTED | GRANTED | REJECTED |
| AGR | 8 | 4 | 7.1 | 4 | 50.0 | 12.5 | 13 | 9 |
| BMED | 19.5 | 8.5 | 17.1 | 10.3 | 62.3 | 43.7 | 12 | 8 |
| CEA | 2.5 | 0 | 2.1 | 0 | 13.6 | 0.0 | 12 | 8 |
| CHE | 21.5 | 8 | 11.5 | 8 | 66.3 | 40.0 | 12 | 9 |
| CHT | 13.5 | 6 | 9 | 4.2 | 50.0 | 57.1 | 13 | 9 |
| CLIM | 17 | 9 | 10.3 | 8.1 | 58.8 | 30.8 | 13 | 9 |
| CSI | 13.5 | 7 | 1.7 | 1.5 | 5.5 | 0.0 | 12 | 9.5 |
| ECO | 5 | 2 | 1.5 | 0 | 0.0 | 0.0 | 12 | 7 |
| ECT | 14 | 6 | 2.6 | 0.8 | 22.9 | 0.0 | 13.5 | 9.3 |
| EDU | 1 | 2 | 0 | 0 | 0.0 | 0.0 | 12 | 9 |
| EEC | 11.5 | 3 | 2.2 | 2 | 10.0 | 0.0 | 12 | 8 |
| ESC | 7 | 4 | 6.5 | 4.9 | 42.9 | 23.2 | 13 | 10 |
| FSB | 10 | 6.5 | 18 | 10.3 | 75.0 | 43.1 | 12 | 8 |
| FST | 16 | 7.5 | 10.6 | 9.1 | 67.7 | 56.3 | 13 | 10 |
| LFF | 10 | 7.5 | 8.8 | 4.9 | 64.7 | 51.3 | 14 | 12 |
| MNA | 10 | 4 | 4.9 | 0.8 | 46.2 | 0.0 | 13 | 7 |
| MST | 19 | 8 | 7.7 | 6.5 | 52.2 | 40.5 | 12 | 8 |
| MTM | 8 | 3.5 | 3 | 1.8 | 14.3 | 0.0 | 12 | 6.5 |
| PHY | 18 | 8 | 10 | 5.5 | 54.8 | 52.9 | 13 | 10 |
| PPH | 11 | 7 | 12.9 | 7.9 | 60.0 | 50.0 | 13 | 10 |
| PSY | 6 | 3 | 3.8 | 0 | 13.3 | 0.0 | 13 | 9 |
| SSC | 2 | 1 | 0 | 0 | 0.0 | 0.0 | 13 | 10 |
| VAB | 10 | 5 | 5.9 | 3.4 | 42.1 | 25.0 | 13 | 9 |
| **ALL AREAS** | **11** | **5** | **7.8** | **3.3** | **50.0** | **16.7** | **13** | **9** |

**Table 5.** Wilcoxon signed-rank test for bibliometric variables (awarded vs. rejected grants)

| Area | OUTPUT | | AV CITATIONS | | %Q1 | | CITATIONS | | Q1 | |
|---|---|---|---|---|---|---|---|---|---|---|
| | Z | ρ | Z | ρ | Z | ρ | Z | ρ | Z | ρ |
| AGR | 590.0 | **2.29E-03** | 533.5 | **4.51E-04** | 518.5 | **2.49E-04** | 463.0 | **4.24E-05** | 366.5 | **6.60E-07** |
| BMED | 375.5 | **2.21E-06** | 471.0 | **8.81E-05** | 505.5 | **2.79E-04** | 240.0 | **3.88E-09** | 284.5 | **3.35E-08** |
| CEA | 93.0 | **6.90E-03** | 105.0 | **1.50E-02** | 95.0 | **3.57E-03** | 97.5 | **7.82E-03** | 92.5 | **2.66E-03** |
| CHE | 1434.0 | **3.67E-11** | 2143.5 | **5.26E-06** | 2299.5 | **3.99E-05** | 1448.5 | **5.00E-11** | 1226.5 | **4.90E-13** |
| CHT | 233.5 | **5.18E-04** | 344.5 | **3.05E-02** | 452.0 | 3.04E-01 | 295.0 | **6.22E-03** | 290.5 | **5.17E-03** |
| CLIM | 49.5 | **4.34E-02** | 65.0 | 1.58E-01 | 31.0 | **5.13E-03** | 49.0 | **4.15E-02** | 36.5 | **1.02E-02** |
| CSI | 406.0 | **1.26E-04** | 623.0 | 6.12E-02 | 710.0 | 2.26E-01 | 493.5 | **2.50E-03** | 686.5 | 1.56E-01 |
| ECO | 1277.5 | **8.97E-03** | 747.0 | **2.47E-08** | 1024.0 | **1.57E-06** | 616.0 | **2.84E-10** | 1005.0 | **7.29E-07** |
| ECT | 262.0 | **8.90E-05** | 361.5 | **5.26E-03** | 322.0 | **1.06E-03** | 295.5 | **4.08E-04** | 262.0 | **6.63E-05** |
| EDU | 1253.5 | 9.53E-01 | 1055.0 | 5.45E-01 | 1057.0 | 5.80E-01 | 1065.0 | 5.88E-01 | 1060.5 | 6.01E-01 |
| EEC | 155.0 | **2.45E-04** | 306.0 | 1.78E-01 | 251.0 | **2.49E-02** | 235.0 | **1.66E-02** | 220.5 | **5.78E-03** |
| ESC | 515.0 | **6.40E-05** | 678.0 | **5.39E-03** | 691.0 | **6.95E-03** | 533.5 | **1.17E-04** | 529.5 | **8.95E-05** |
| FSB | 6654.5 | **2.58E-05** | 5196.5 | **5.10E-10** | 5662.5 | **2.41E-08** | 4608.0 | **1.97E-12** | 4762.0 | **8.00E-12** |
| FST | 480.0 | **2.54E-05** | 779.5 | 5.69E-02 | 744.5 | **3.07E-02** | 481.0 | **2.67E-05** | 468.0 | **1.62E-05** |
| LFF | 313.5 | 5.07E-02 | 156.5 | **2.47E-05** | 307.0 | **4.10E-02** | 198.0 | **3.15E-04** | 268.5 | **9.66E-03** |
| MNA | 116.5 | **3.93E-04** | 108.0 | **2.08E-04** | 110.5 | **1.98E-04** | 99.5 | **1.06E-04** | 105.5 | **1.29E-04** |
| MST | 583.5 | **3.71E-05** | 933.5 | 6.27E-02 | 913.5 | **4.72E-02** | 645.0 | **2.05E-04** | 634.0 | **1.49E-04** |
| MTM | 441.0 | **9.89E-05** | 637.5 | **1.52E-02** | 662.0 | **2.20E-02** | 552.0 | **2.26E-03** | 564.0 | **2.26E-03** |
| PHY | 493.0 | **4.44E-05** | 645.5 | **1.86E-03** | 1338.0 | 9.57E-01 | 517.0 | **8.58E-05** | 641.5 | **1.68E-03** |
| PPH | 882.0 | **2.55E-04** | 789.5 | **3.04E-05** | 1056.0 | **7.05E-03** | 644.5 | **5.96E-07** | 743.5 | **8.85E-06** |
| PSY | 994.0 | **2.74E-04** | 785.5 | **1.16E-06** | 1010.5 | **8.48E-05** | 747.5 | **3.80E-07** | 954.0 | **1.69E-05** |
| SSC | 1213.0 | 6.26E-02 | 1157.0 | **1.60E-03** | 1389.5 | 1.62E-01 | 1163.5 | **1.97E-03** | 1392.0 | 1.72E-01 |
| VAB | 972.0 | **1.43E-05** | 1005.5 | **3.09E-05** | 1104.0 | **2.26E-04** | 847.5 | **7.19E-07** | 843.5 | **5.50E-07** |

**Z:** Wilcoxon-test value; **ρ:** ρ-value. In bold: Statistically significant differences (p<0.05)

**Table 6.** Pearson's and Spearman's correlation coefficient between bibliometric indicators and PIs' CV ratings by research fields

| Area | Pearson | | | | | Spearman | | | | |
|---|---|---|---|---|---|---|---|---|---|---|
| | OUTPUT | AV CITATIONS | %Q1 | CITATIONS | Q1 | OUTPUT | AV CITATIONS | %Q1 | CITATIONS | Q1 |
| AGR | **0.40** | 0.25 | **0.32** | **0.37** | **0.46** | **0.49** | **0.40** | 0.34 | **0.53** | **0.57** |
| BMED | **0.37** | 0.20 | **0.35** | **0.32** | **0.45** | **0.51** | **0.40** | **0.37** | **0.61** | **0.59** |
| CEA | **0.46** | 0.21 | **0.34** | **0.35** | **0.40** | **0.45** | 0.32 | **0.33** | **0.37** | **0.37** |
| CHE | 0.56 | **0.42** | **0.34** | **0.44** | 0.59 | **0.68** | **0.43** | **0.32** | **0.67** | **0.69** |
| CHT | **0.45** | 0.07 | **0.31** | **0.25** | **0.43** | **0.50** | 0.23 | **0.27** | **0.42** | **0.53** |
| CLIM | **0.55** | **0.42** | **0.60** | **0.54** | **0.62** | **0.59** | **0.48** | **0.62** | **0.60** | **0.69** |
| CSI | **0.45** | 0.19 | 0.11 | **0.27** | 0.21 | **0.48** | 0.24 | **0.27** | **0.40** | **0.32** |
| ECO | **0.29** | **0.40** | **0.35** | **0.49** | **0.41** | **0.34** | **0.54** | **0.42** | **0.61** | **0.44** |
| ECT | **0.48** | 0.01 | **0.28** | **0.32** | **0.37** | **0.52** | **0.28** | **0.32** | **0.43** | **0.49** |
| EDU | -0.04 | 0.10 | 0.11 | -0.03 | -0.01 | -0.06 | 0.16 | 0.05 | 0.14 | 0.04 |
| EEC | **0.63** | 0.18 | **0.28** | **0.46** | **0.58** | **0.73** | **0.37** | **0.51** | **0.65** | **0.69** |
| ESC | **0.37** | 0.29 | 0.13 | **0.25** | **0.30** | **0.49** | **0.32** | 0.16 | **0.51** | **0.42** |
| FSB | **0.33** | 0.17 | **0.30** | **0.36** | 0.46 | **0.39** | **0.31** | **0.28** | **0.49** | **0.53** |
| FST | **0.58** | 0.25 | **0.30** | **0.54** | **0.55** | **0.62** | 0.25 | 0.20 | **0.60** | **0.59** |
| LFF | **0.47** | 0.29 | 0.13 | **0.55** | **0.47** | **0.50** | **0.50** | 0.05 | **0.68** | **0.46** |
| MNA | **0.60** | **0.53** | **0.56** | **0.51** | **0.58** | **0.68** | **0.66** | **0.62** | **0.75** | **0.72** |
| MST | **0.52** | 0.22 | **0.38** | **0.29** | **0.50** | **0.63** | 0.23 | 0.40 | **0.54** | **0.63** |
| MTM | **0.52** | **0.21** | **0.32** | **0.48** | **0.50** | **0.57** | **0.40** | **0.34** | **0.55** | **0.48** |
| PHY | **0.37** | **0.40** | -0.07 | **0.34** | 0.31 | **0.48** | **0.47** | 0.00 | **0.56** | **0.46** |
| PPH | **0.30** | **0.37** | **0.23** | **0.48** | **0.49** | **0.46** | **0.41** | **0.26** | **0.56** | **0.54** |
| PSY | **0.42** | **0.53** | **0.38** | **0.42** | **0.45** | **0.48** | **0.64** | **0.46** | **0.65** | **0.51** |
| SSC | 0.18 | **0.27** | 0.17 | 0.18 | 0.09 | **0.20** | **0.26** | 0.14 | **0.25** | 0.13 |
| VAB | **0.41** | **0.29** | **0.29** | **0.47** | **0.40** | **0.61** | **0.48** | **0.35** | **0.66** | **0.58** |

In bold: Statistically significant differences (p<0.05)

**Table 7.** Stepwise logistic regression analysis by area. Explanatory variables for granting proposals, odds ratio and goodness of fit measures and prediction ability measures.

| Area | G2 | gl | ρ | AUC | CCR | Explanatory variables and odds ratios | Intercept |
|---|---|---|---|---|---|---|---|
| AGR | 91.02 | 84 | 0.28 | 0.80 | 72.09% | Q1=1.59 | I=0.36 (2.74) |
| BMED | 74.33 | 82 | 0.71 | 0.88 | 82.56% | Q1=1.57; OUTPUT=0.87; AV CITATIONS=1 | I=0.11 (1.15) |
| CEA | 39.95 | 35 | 0.269 | 0.73 | 70.27% | Q1=3.50 | I=0.46 (2) |
| CHE | 175.47 | 185 | 0.68 | 0,83 | 71.66% | Q1=1.20 | I=0.52 (2) |
| CHT | 72.976 | 73 | 0.48 | 0.76 | 66.67% | OUTPUT= 1.11 | I=1 |
| CLIM | 19.50 | 28 | 0.88 | 0.89 | 81.25% | Q1=1.22; %Q1=1.08; AV CITATIONS=1 | I=0.01 (100) |
| CSI | 89.21 | 77 | 0.14 | 0.79 | 73.75% | OUTPUT= 1.21; Q1=0.44; %Q1=1 | I= 0.22 (4) |
| ECO | 114.00 | 114 | 0.48 | 0.82 | 78.63% | CITATIONS=2.67; Q1=1 | I=0.30 (3.3) |
| ECT | 66.97 | 68 | 0.51 | 0.80 | 72.22% | OUTPUT=1.17; %Q1=1.06; CITATIONS=1 | I=0.17 (6) |
| EDU | 123.14 | 91 | 0.01 | 0.60 | 50.54% | | |
| EEC | 53.73 | 53 | 0.456 | 0.83 | 78.95% | OUTPUT=1.34; %Q1=1; CITATIONS=0.98 | I=0.29 (3.3) |
| ESC | 102.17 | 95 | 0.29 | 0.74 | 65.98% | Q1=1.53 | I=1 |
| FSB | 289.72 | 310 | 0.79 | 0.80 | 68.79% | Q1=1.44; OUTPUT=0.83; CITATIONS=1 | I=1 |
| FST | 102.20 | 87 | 0.13 | 0,77 | 66.67% | OUTPUT= 1.10; %Q1=1 | I= 0.13 (8.3) |
| LFF | 58.81 | 57 | 0.419 | 0.81 | 69.49% | AV CITATIONS=1.51 | I=0.09 (10) |
| MNA | 42.79 | 47 | 0.658 | 0.87 | 78% | OUTPUT=1.18; %Q1=1.04 | I=0.23 (4) |
| MST | 113.32 | 105 | 0.27 | 0,75 | 67.29% | OUTPUT= 1.08 | I=1 |
| MTM | 93.39 | 102 | 0.72 | 0.76 | 68.57% | OUTPUT= 1.18; %Q1=1 | I=1 |
| PHY | 89.53 | 120 | 0.98 | 0.81 | 71.77% | OUTPUT=1.08; %Q1=1; AV CITATIONS=1 | I=1 |
| PPH | 116.67 | 115 | 0.44 | 0.77 | 70.34% | OUTPUT=0.90; CITATIONS=1.02 | I=1 |
| PSY | 130.77 | 109 | 0.08 | 0.76 | 72.57% | Q1=1; AV CITATIONS=1; OUTPUT=1 | I=0.32 (3.3) |
| SSC | 137.96 | 105 | 0.02 | 0.58 | 61.11% | | |
| VAB | 139.34 | 124 | 0.16 | 0.76 | 65,87% | Q1=1.33 | I=1 |

The logistic regression model does not fit for p-values > 0.05